\documentclass{aastex631}
\graphicspath{{./}{figures/}}

\shorttitle{Inferring Solar Flows Using Bayesian Methods}
\shortauthors{Herczeg \& Jackiewicz}

\usepackage{amsmath}
\usepackage[figuresright]{rotating}
\usepackage{tabularx}
\usepackage{layouts}
\usepackage{graphicx}

\begin{document}
\title{Inferring the Solar Meridional Circulation Flow Profile By Applying Bayesian Methods to Time-Distance Helioseismology}
\author{Aleczander Herczeg}
\author{Jason Jackiewicz}
\affiliation{Department of Astronomy, New Mexico State University, Las Cruces, NM 88003, USA}

\begin{abstract}

Mapping the large-scale subsurface plasma flow profile within the Sun has been attempted using various methods for several decades. One such flow in particular is the meridional circulation, for which numerous studies have been published. However, such studies often show disagreement in structure. In an effort to constrain the flow profile from the data, a Bayesian Markov chain Monte Carlo framework has been developed to take advantage of the advances in computing power that allow for the efficient exploration of high-dimensional parameter spaces. This study utilizes helioseismic travel-time difference data covering a span of twenty-one years and a parametrized model of the meridional circulation to find the most likely flow profiles. Tests were carried out on artificial data to determine the ability of this method to recover expected solar-like flow profiles as well as a few extreme cases. We find the method capable of recovering the input flows of both single- and double-cell flow structures. Some inversion results indicate potential differences in meridional circulation between the two solar cycles in terms of both magnitude and morphology, in particular in the mid convection zone. Of these, the most likely solutions show that solar cycle 23 has a large, single-celled profile, while cycle 24 shows weaker flows in general and hints towards a double-celled structure. 
\end{abstract}

\section{Introduction} \label{sec:intro}

Helioseismology is a collection of methods to study the subsurface structure of the Sun from oscillation properties measured at the surface. One such method, known as time-distance helioseismology \citep{1993Natur.362..430D}, probes the solar interior by measuring the time it takes for acoustic wave packets to travel from one point (point A) on the surface to another (point B) on its path through the solar interior. These waves are sensitive to the plasma flows as they propagate through the interior, shortening the travel time when the flow is aligned to the direction of propagation and lengthening the travel time when antialigned. Taking the difference between the travel times from point A to point B and the travel times from point B to point A yields a travel-time shift for that particular point pair. With a sufficient number of travel time shifts measured at different point separation distances (hereafter referred to as skip distances) and latitudes, a map of travel-time shifts can be created. Then, by using a function that relates the flow field to the travel-time differences, the flows can be inverted using the travel-time difference map. For this work, the focus of these measurements will be on understanding the meridional circulation flow profile. Meridional circulation is thought to be a flow that moves plasma from the equator to the poles at the surface, with a return flow from the poles to the equator at some depth below the surface. This flow is difficult to characterize - the speed of the flow at the surface is only on the order of 15 to 20 meters per second, significantly slower than typical flows on the solar surface such as granulation. The subsurface meridional circulation is particularly difficult to characterize, as it is even smaller in magnitude and hidden by realization noise \citep{2008ApJ...689L.161B}.

Inversion methods, such as the subtractive optimally localized averages (SOLA) method \citep{1994A&A...281..231P}, have been employed for decades in helioseismology \citep[in works such as][]{1998ApJ...505..390S, 2003ARA&A..41..599T, 2009LRSP....6....1H, 2015ApJ...805..133J}. One advantage to this method is that a parametric model of the flow profile is not needed. As long as the flow is linear, there are no additional assumptions needed. A disadvantage is that only one primary solution can be found, with minor variations available by tweaking the  trade-off parameters.
Other methods make use of a regularized least squares (RLS) technique, which do utilize parametric flow models as discussed in papers such as \citet{2015ApJ...813..114R}. However, these methods also suffer from requiring somewhat arbitrary choices of regularization (smoothing) parameters to yield acceptable solutions. As helioseismic inverse problems are not well posed, such smoothing is necessary.


A Markov chain Monte Carlo (MCMC) Bayesian inference method is a tool to robustly explore  a model's parameter space. Methods of this type for helioseismic inversions have already shown promise in preliminary studies \citep{2020SoPh..295..137J}.
Using Bayes's theory, each parameter value is assigned a probability indicating its likelihood  given the data set. Therefore, rather than the output of the inversion being the most likely parameters, the output is a multi-dimensional posterior probability density function (PDF). While this method yields a great deal of information for each model parameter, the downside is that the forward calculations must be performed many times to effectively sample the parameter space, with the number of computations increasing significantly for each parameter added to the model and each data point added to the data set. Additionally, a flow model is used, and care must be taken to select a model that can reproduce the flows with adequate detail without being overly restrictive.

In this work, we apply this technique in the framework of time-distance helioseismology to study the meridional circulation profile in the solar convection zone. The primary motivation for doing so is due to the lack of a consensus among previous studies of this flow. Recent inferences of meridional circulation  \citep[][to name just a few]{2013ApJ...774L..29Z,2013ApJ...778L..38S,2015ApJ...805..133J,2015ApJ...813..114R,2018A&A...619A..99L,2020Sci...368.1469G} are not in agreement. One outstanding question regarding the meridional flow profile is whether or not the flow is a single or multiple cell in depth. For example, there is evidence for a double-celled profile from the \cite{2013ApJ...774L..29Z} analysis and single-celled profiles from the \cite{2015ApJ...805..133J}, \cite{2015ApJ...813..114R}, and \cite{2020Sci...368.1469G} analyses. 

The inconsistencies in these studies are certainly partly due to differences in the methods used to compute each set of travel-time differences. Likely more important are the effects of systematic errors in helioseismic data from different instruments \citep{2020Sci...368.1469G}.

Here we apply Bayesian inferences to two published helioseismic data sets over the past two solar cycles made publicly available by \cite{2018A&A...619A..99L} and \cite{2020Sci...368.1469G}. Details of the data sets will be discussed in Section\,\ref{sec:data}. \citet{2018A&A...619A..99L} also describe the model for the meridional circulation which we adopt. We do not attempt to reproduce any of the measurements or systematic-error corrections from these published data - such an effort is beyond the scope of this work. Instead, the goal is to use the probabilistic inversion method to explore the resulting meridional flows and compare to other inversion approaches.





Inversions require sensitivity kernels that relate flow perturbations along a given ray path through the solar interior to travel-time perturbations. In this study, the kernels used are computed following the ray approximation \citep{1996ApJ...461L..55K, 1997ASSL..225..241K}. Ray kernels were chosen for this work rather than Born kernels due to their similar accuracy in weak flow regimes such as expected for meridional circulation \citep{2004ApJ...616.1261B}. 
An overview of the data set used and the steps taken to prepare the data for use with the Bayesian methods is described in Section \ref{sec:data}. Specifics of the forward model, the Bayesian and MCMC techniques, and the details of the Bayesian parameters are next discussed in Section \ref{sec:methods}. Results of this new method and comparisons to traditional techniques are laid out in Section \ref{sec:results}, followed by a final summary in Section \ref{sec:summary}.

\section{Data} \label{sec:data}

Two sets of data were analyzed in this study, with both sets spanning solar cycles 23 and 24, travel-time skip distances from \(6^\circ\) to \(42^\circ\), and latitude coverage up to \(50^\circ\) north and south. The first travel-time difference data set is described in \cite{2018A&A...619A..99L} and made publicly available online\footnote{\url{http://cdsarc.u-strasbg.fr/viz-bin/qcat?J/A+A/619/A99}}. The measurements were computed from Doppler data that span from 1996 to 2017, with 12 years of data from solar cycle 23 (1996 to 2008) and 9 years of data from solar cycle 24 (2008 to 2017). Though the entirety of cycle 24 was not captured in this data set, the two sets will be referred to as cycle 23 and cycle 24, respectively. Such a time span was made possible by combining data from both the Michelson Doppler Imager (MDI) \citep{1995SoPh..162..129S} on board the Solar and Heliospheric Orbiter and the Helioseismic and Magnetic Imager (HMI) \citep{2012SoPh..275....3P} on board the Solar Dynamics Observatory. To allow for the use of both instruments, \cite{2018A&A...619A..99L} removed systematic errors in the travel time data such that the period of overlap between the two instruments yielded matching travel time differences. One significant source of systematic error in helioseismology is the center-to-limb effect. While the physical nature of the center-to-limb effect is still unknown \citep{2012ApJ...749L...5Z,2012ApJ...760L...1B,2018A&A...617A.111S}, its measurement and removal significantly improves the match between MDI and HMI measurements. The processed travel times show a substantial asymmetry between the northern and southern hemispheres in both short- and long-term averaged data. The asymmetry manifests in both solar cycles, but is stronger for cycle 24. Beyond skip distances of around \(12^\circ\), the northern hemisphere south-north travel times are consistently and significantly lower than the southern hemisphere travel times. At large skip distances (greater than \(30^\circ\) or so), the northern hemisphere travel time differences drop to zero. While the effect is less significant in cycle 23, it is still present. 

Many additional steps were taken by \citet{2018A&A...619A..99L} in order to eliminate as many other systematic errors as possible. For the MDI portion of the time span studied, data were only selected when the P-angle was zero, and the data were manually inspected to remove corrupted files. Additionally, a one hour running mean was subtracted from the MDI data to account for large scale flow effects and a bandpass filter was applied to remove effects of granulation, supergranulation, and propagating flows above the cutoff frequency. To match the HMI and MDI data sets, the MDI data's P-angle offset was adjusted to match HMI according to a known instrumental difference. Data from both instruments are masked to remove effects of strong magnetic field, where the cutoff is determined by 5 times the standard deviation in magnetic field strength for each instrument. For the HMI data, an additional P-angle adjustment has been made due to a discovered difference between measured and true P-angle during transits.


The second data set is also publicly available\footnote{\url{https://edmond.mpdl.mpg.de/dataset.xhtml?persistentId=doi:10.17617/3.NAPBUA&version=1.0}}, published in \cite{2020Sci...368.1469G}. MDI P-angle adjustments, areas of strong magnetic field, and the center-to-limb effect were all accounted for in order to minimize systematic errors prior to computing travel time measurements. However, HMI data were considered unreliable from unknown systematics and were discarded in their analysis. These data are split into two sets of travel-time difference measurements: cycle 23, which spans from May 1st, 1996 until 2008, and cycle 24, which spans from 2008 until April 30th, 2019. Data up until April 30th, 2003 come from MDI, while all data after come from the Global Oscillation Network Group (GONG).

To summarize, one of the data sets we study includes MDI and HMI data, while the other comprises MDI and GONG, and therefore direct comparisons between the inversions should be viewed cautiously due to unknown instrument-dependent systematics that might be unaccounted for. It is still worthwhile to utilize HMI data, particularly given its long time series. For example, the recent detection and characterization of solar Rossby waves in HMI measurements \citep{2018NatAs...2..568L,2021A&A...652A..96M} has been reproduced from other sources, such as GONG \citep{2020A&A...635A.109H}.



Finally, while the publicly available data used for this analysis cover a substantial amount of time, it is noted by \cite{2020Sci...368.1469G} that directly detecting the deepest meridional flows near $0.7R_\odot$ would require somewhere on the order of 100 years of data, which is significantly more than is currently available. As such, the resulting flows at the  bottom of the convection zone are less reliable than the flows near the surface.

\section{Methods} \label{sec:methods}
\subsection{Bayesian Inference} \label{subsec:bayes}
Bayesian inference is a probabilistic approach to data analysis. A key advantage of this type of inference in general is the ability to compute a posterior probability distribution function (PDF) of model parameters, where typical methods would merely provide the maximum of the posterior PDF. Analyzing the full PDF for a parameter can reveal interesting phenomena, such as additional local maxima beyond the most likely parameter value that would be hidden without having knowledge of the PDF. However, sampling such a PDF can be very computationally expensive, especially in models with even a moderate number of free parameters. Algorithms have been developed that efficiently allow us to sample this distribution, such as the Markov chain Monte Carlo (MCMC) method.

The concept of an MCMC method encapsulates two separate ideas. First is the idea of a Monte Carlo method. A Monte Carlo method is a type of algorithm that seeks to utilize random sampling to either draw a uniform sample of some probability distribution or scatter points within a domain for numerical integration or optimization. This concept is coupled with a Markov chain, which is any sort of model that describes a string of events in which the probability of any event is only reliant on the event or state that came directly before it. Combining a Monte Carlo algorithm with a Markov chain yields the MCMC method. This combination takes advantage of the the random sampling offered by Monte Carlo methods with the ability to prescribe a Markov chain stepping algorithm or probability distribution function with which to modify the random sampling. Hence, one can create an ensemble of "walkers" that start at some arbitrary location in parameter space and step or "walk" to new values of each parameter based on the location the step is going from. In this work, it was found that using 36 walkers for each of the 12 parameters provided a balance between parameter convergence and run time. Parameter values prefer to move away from locations in parameter space that lower the likelihood of the fit. In this way, the algorithm explores the parameter space more densely in regions of high likelihood. To quantify the likelihood of a model realization, a multivariate normal distribution was chosen.

One particularly useful aspect of Bayesian statistics is the inclusion of a prior probability distribution. The Bayesian framework includes an additional term that is associated with the expected likelihood of a certain outcome before even performing any analysis. Selecting very narrow priors for parameters can restrict the results to a narrow range of possible solutions, so care must be taken in selecting priors. In many cases, the ability to use prior knowledge can be very useful. For an example in the context of solar flows, a prior constraining the surface meridional circulation flow to values near those measured using other methods can be introduced to avoid wasting computation time on flow models where the flow is in wild disagreement with well-constrained observational evidence. In this work, the prior ranges are left relatively wide to allow for a large number of possible solutions, while constraining them enough to prevent wildly nonphysical results from being computed. Much akin to setting smoothing parameters in other methods, it should be noted that prior selection is an inexact science, and terms like "relatively wide" or "narrow" are more a matter of subjective taste than rigorously tested and verified definitions.

An efficient and effective implementation of these Bayesian MCMC methods is made possible by the \texttt{emcee} package developed for Python \citep{2013PASP..125..306F}. In this work, the \texttt{emcee} StretchMove was utilized as a stepping algorithm.

\subsection{Flow Model} \label{subsec:flowmodel}
In order to interpret the travel-time difference measurements, a forward model of the subsurface flows in question is required. As developing a new meridional circulation flow model is beyond the scope of this work, we used a slightly modified version of the model created and described in \cite{2018A&A...619A..99L}. The model flows are computed for two separate regimes and then summed to produce the full flow. The first regime is the main meridional circulation (MC) flow. The second regime can model the near-surface inflow around magnetically active latitudes, called the local cellular (LC) flow. It is well known that active regions have inflows that work to reduce the poleward velocity of the near-surface meridional circulation, as measured using helioseismic methods \citep{2001IAUS..203..189G} and independent correlation tracking \cite{2017A&A...606A..28L}. While the MC and LC portions of the model are separate and an MC-only inversion can be performed, we found that including the LC component led to a better fit to the solar data in all cases. While this LC flow varies in strength and central latitude throughout a solar cycle \citep{2001ApJ...559L.175C}, the impact of the average LC flow is clearly visible in the averaged near-surface travel time measurements. As such, models including LC flow yield a better fit to the data than models that only utilize MC flow. With \(\theta\) representing colatitude, the total flow profile is of the form \(\mathbf{u} = u_{r}(r,\theta) \mathbf{\hat{r}} + u_{\theta}(r,\theta) \boldsymbol{\hat{\theta}}\), with radial and horizontal elements defined in a separable form as

\begin{equation} \label{eq:ur}
u_r(r,\theta) = f(r) g(\theta),
\end{equation}
\begin{equation} \label{eq:uth}
u_\theta(r,\theta) = -F(r) G(\theta).
\end{equation}

The flow field conserves mass throughout the model domain by using $\nabla \cdot (\rho \mathbf{u})=0$, yielding the expressions

\begin{equation} \label{eq:Fr}
  F(r) = \frac{1}{r\rho(r)}\frac{d}{dr} \Big(r^2\rho(r)f(r) \Big),
\end{equation}
\begin{equation} \label{eq:gth}
  g(\theta) = \frac{1}{\sin\theta}\frac{d}{d\theta} \Big(\sin\theta G(\theta) \Big),
\end{equation} where \(\rho\) is the density as a function of radius, which we take from Model S \citep{1996Sci...272.1286C}. Finally, the latitudinal dependence of the horizontal flows is defined as

\begin{equation} \label{eq:Gth}
  G(\theta) = 
  \left\{
  \begin{array}{ll}
    \text{sin}\big(2\pi\frac{\theta-\theta_n}{\theta_s-\theta_n}\big) \text{sin}^\alpha \big(\pi\frac{\theta-\theta_n}{\theta_s-\theta_n}\big) & \theta_n \leq \theta \leq \theta_s, \\
    0 & \text{otherwise}.
  \end{array}
  \right.
\end{equation} The \(\theta_s\) and \(\theta_n\) terms refer to the southern and northern latitude limits respectively. While both the MC and LC flows are computed using these same equations, different parameters control the generation of each flow. For the MC case, the latitude boundaries are simply the poles. On the other hand, the LC boundaries require a bit more care. Two additional free parameters are introduced to handle these boundaries, "width" and "center". The "width" parameter sets the latitudinal extent of each cell, while the "center" parameter sets the midpoint of each cell. Often, the two cells overlap, and the resulting LC flow profile is a mixture of these two cells around the equator. For the purposes of this work, the local cells are treated as being symmetrical about the equator. In principle, these two parameters could be duplicated and adjusted independently to allow for asymmetry in the LC component, though this is not considered here. The \(\alpha\) term describes the skewness of the flow profile.

\begin{table}
  \centering
  \begin{tabular*}{\textwidth}{c @{\extracolsep{\fill}} cc}
    \hline
    Parameter & Description & Prior Range \\ [0.5ex]
    \hline\hline
    MC\textsubscript{1} & \(F(r = 0.75R_\odot)\) from Eq. \ref{eq:Fr} & -100 to 100 (ms\(^{-1}\))\\
    MC\textsubscript{2} & \(F(r = 0.80R_\odot)\) from Eq. \ref{eq:Fr} & -60 to 60 (ms\(^{-1}\)) \\
    MC\textsubscript{3} & \(F(r = 0.85R_\odot)\) from Eq. \ref{eq:Fr} & -60 to 60 (ms\(^{-1}\)) \\
    MC\textsubscript{4} & \(F(r = 0.90R_\odot)\) from Eq. \ref{eq:Fr} & -60 to 60 (ms\(^{-1}\)) \\
    MC\textsubscript{5} & \(F(r = 0.95R_\odot)\) from Eq. \ref{eq:Fr} & -60 to 60 (ms\(^{-1}\)) \\
    MC\textsubscript{6} & \(F(r = 1.00R_\odot)\) from Eq. \ref{eq:Fr} & -60 to 100 (ms\(^{-1}\)) \\
    MC\textsubscript{7} & Skewness of MC Flow (\(\alpha\) in Eq. \ref{eq:Gth}) & 1 to 3 \\
    \hline
    LC\textsubscript{1} & Width of LC Flow & 0 to 180 (deg)\\
    LC\textsubscript{2} & Central Latitude of LC Flow & 0 to 25 (deg)\\
    LC\textsubscript{3} & Rate of LC Flow Decrease with Depth & 50 to 1000 (ms\(^{-1}\)R\(_\odot\)\(^{-1}\))\\
    LC\textsubscript{4} & Surface Velocity of LC Flow & 1 to 45 (ms\(^{-1}\))\\
    LC\textsubscript{5} & Skewness of LC Flow (\(\alpha\) in Eq. \ref{eq:Gth}) & 2 to 10 \\
    \hline
  \end{tabular*}
  \caption{Description and prior ranges for each of the twelve parameters used in the forward model: seven for the MC component of the flow profile, and five for the LC component.}
  \label{tab:paramnames}
\end{table}

The radius points and associated flow values selected differ for the MC and LC cases, and are determined independently. For the MC case, the lower boundary is set to 0.7\(R_\odot\) and the upper boundary to 1.0\(R_\odot\). The \(F(r)\) values at all points other than the lower boundary are free parameters in the model. The horizontal velocity at the lower boundary is adjusted until \(f(r)\) vanishes at the boundaries as described above. To create the LC flow profile, two parameters are used that are not described in the original model by \cite{2018A&A...619A..99L}. The peak horizontal flow velocity (at the surface) is set by the LC\(_4\) parameter. The next key element of the LC flow profile is the depth at which the horizontal flow changes sign. Instead of setting this depth as a parameter itself, we instead define a parameter (LC\(_3\)) which describes the rate at which the velocity decreases from the peak surface value. Using the peak surface velocity and the slope parameter, the depth at which the flow velocity reaches zero and changes directions is trivial to determine. Contrary to the MC flow case, the LC flow's lower boundary is not at the tachocline, but instead must be computed. For a given value of the lower boundary, the flow is calculated using spline interpolation between the upper and lower boundaries using the depth at which the flow changes sign and an additional point halfway between this depth and the surface as reference points for the interpolation. Using Eq. \ref{eq:Fr}, the vertical flow \(f(r)\) can be solved for. Then, a root finding algorithm is used to adjust the lower boundary location. The root finder iterates until a maximum depth is found at which the vertical flow at said boundary is equal to zero.
Free parameters of the model are listed in Table \ref{tab:paramnames}. The parameters are split into two categories - MC and LC - to more easily distinguish the parameters of the two independent flow profiles. While the model as written uses the convention of positive flow values indicating southward flow, our work presents positive flows as being northward in all cases.

Note in particular that the MC\textsubscript{1} through MC\textsubscript{6} parameters are used to compute the horizontal flow at a given depth, but also vary depending on the vertical velocity and latitude. As an example, taking the MC\textsubscript{6} value for "Liang 23" in Table\,\ref{tab:parranges} of 18.12 and computing the horizontal flow at a latitude of \(40.4^\circ\) north yields a velocity of 10.46 meters per second northward.

\begin{figure}
  \centering
  \includegraphics[width=\textwidth]{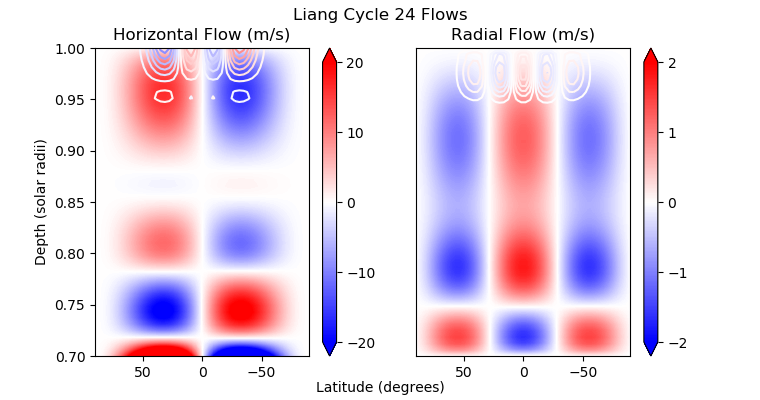}
  \caption{Full illustration of the flow profiles produced using both the data and model from \cite{2018A&A...619A..99L}. This model is computed using the most likely values from our inversion of the cycle 24 data (column two of Table\,\ref{tab:parranges}). The filled contours represent the MC component, and the contour lines represent the LC component. Note the color scale in the left panel is restricted such that the large near-tachocline values are saturated to better illustrate the detail above a depth of \(0.75_odot\). The color scale is shared for the MC and LC components within each panel. The contour lines represent 12 and 10 evenly spaced levels between the minimum and maximum velocities for the LC component in the left and right panels respectively. (For the horizontal LC flow, the LC flow ranges from -9.96 to 9.96 meters per second, while the radial LC flow ranges from -0.26 to 0.23 meters per second.)}
  \label{fig:flowplots}
\end{figure}

To illustrate the full flow profile, the flow profile using the most likely parameter values from our inversion of the \cite{2018A&A...619A..99L} cycle 24 data (see Table\,\ref{tab:parranges}) is shown in Figure\,\ref{fig:flowplots}. Here, one can see the relative contributions and morphologies of the MC and LC components of the flow for both the horizontal and radial components.

\subsection{Inversion Tests with Artificial Data} \label{subsec:recovery}

\begin{figure}
  \centering
  \includegraphics[width=.7\textwidth]{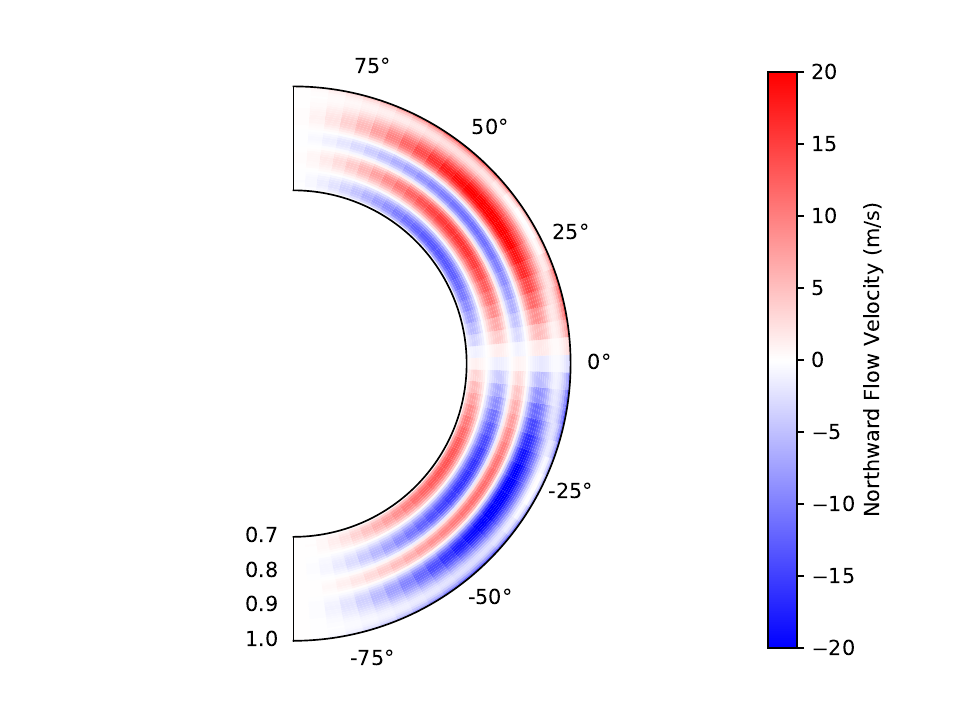}
  \caption{Artificial flow profile used to test the ability to recover highly multi-cellular flow profiles in the ray approximation. This flow profile was used to forward model travel-time difference data, to which varying levels of noise were added. This noisy travel-time difference data was then used as the input to the analysis to test the accuracy of the recovered flows.}
  \label{fig:polar_multicell}
\end{figure}

The inversion uses a likelihood function that compares the measurements to forward model travel-time differences. The forward model uses ray approximation kernels 
computed following the formalism outlined in \cite{2015ApJ...805..133J}. To understand the sensitivity of ray kernels to potential sharp changes in flow velocity with depth, tests were carried out to recover a highly multi-cellular flow profile from artificial travel times with varying levels of noise. The artificial flow profile used to create the travel-time differences used as the input for the recovery is shown in Figure \ref{fig:polar_multicell}. This flow was then used to forward-model travel-time differences to which Gaussian noise was added. Various levels of noise were tested by adjusting the standard deviation of the distribution such that the resulting signal-to-noise ratio of the artificial data ranged from 2 to 20. While the true errors vary with latitude and skip distance, we consider a uniform error distribution to be valid for testing in this case as the noise levels tested here are consistently higher than the noise found in the data.

\begin{figure}
   \centering
   \includegraphics[width=0.7\textwidth]{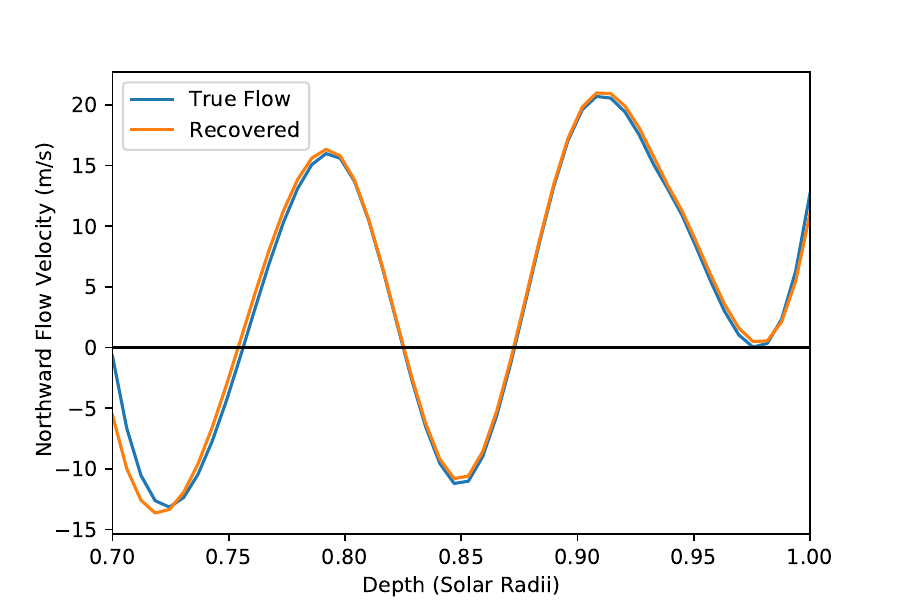}
   \caption{Recovery of the artificial data set to test the efficacy of ray kernels in quickly changing flow regimes. Northward horizontal flow velocity is displayed as a function of depth for an average of four latitude points centered at \(30^\circ\).}
   \label{fig:artificialcomp}
 \end{figure}

 Recovery of this test data set was successful with ray kernels. The depths at which the flows change sign were accurate to within a couple percent of a solar radius. The largest mismatch occurred at the boundaries - most notably, the bottom boundary, where the model has no free parameter for the flow velocity as discussed in Section \ref{subsec:flowmodel}. A comparison between the true flow profile and the best-fit recovery as a function of depth at a latitude of \(30^\circ\) is shown in the top panel of Figure\,\ref{fig:artificialcomp}.

While the model is simple (especially in its lack of parameters controlling latitudinal variations), it is sufficiently robust to capture radial variations in the horizontal flow velocity, which is the primary concern of this work.


\subsection{Inversions with Solar Data} \label{subsec:realinv}

To properly adjust prior ranges and the number of iterations required to converge to a well-fitting solution, a number of inversions were computed to settle on acceptable values for each of these variables. Priors for each flow parameter were left relatively wide so as to not unduly influence the resulting flow profile. By observing the autocorrelation of the parameters of each inversion run, it was determined that 300,000-600,000 iterations were sufficient for each parameter to converge.

While runs performed on the full, un-averaged data set are quite computationally expensive, testing revealed that binning the data reduced the quality of the fit too much to be a reasonable trade-off with computation time. As such, the only alteration made to the data set is an iterative sigma-cutting to remove the most egregious outlier points. Travel-time differences with values more than 3 sigma from the average are removed from the data set, then a new average and sigma are computed and the process is repeated until no points are removed. This results in, on average, about 2 to 3 percent of the points being cut.

Following the same procedure as the \cite{2018A&A...619A..99L} data, an analysis was performed on the data made publicly available by \cite{2020Sci...368.1469G}. This investigation differs from the previous investigation in that there are published non-Bayesian inversions performed on this data set to compare the Bayesian approach to. Beyond the direct comparison of this work's Bayesian inversions to \cite{2020Sci...368.1469G}'s non-Bayesian inversions, utilizing a second data set allows for a more robust exploration of the model's ability to produce realistic flow profile predictions using real solar data.

Errors for this data set were provided in the form of a full matrix of the covariance between each pair of points. The diagonal of the two-dimensional covariance matrix corresponds to the variance on each individual point, while the off-diagonal components (x,y) provide the joint variability of the points x and y together. After testing, it was found that restricting the errors to only the diagonals of the matrix (analogous to the errors used in the \cite{2018A&A...619A..99L} data sets) yielded a slightly inferior fit to the data.

Two Bayesian inversions were performed on the \cite{2018A&A...619A..99L} data set; one for cycle 23 and one for cycle 24. For the \cite{2020Sci...368.1469G} data set, four inversions were performed: for each solar cycle, an inversion was performed using the full error covariance matrix as well as an inversion using the diagonal components of the error covariance to facilitate a more direct comparison between the results of the two data sets.

\section{Results and Discussion} \label{sec:results}

\begin{figure}
  \centering
  \includegraphics[width=\textwidth]{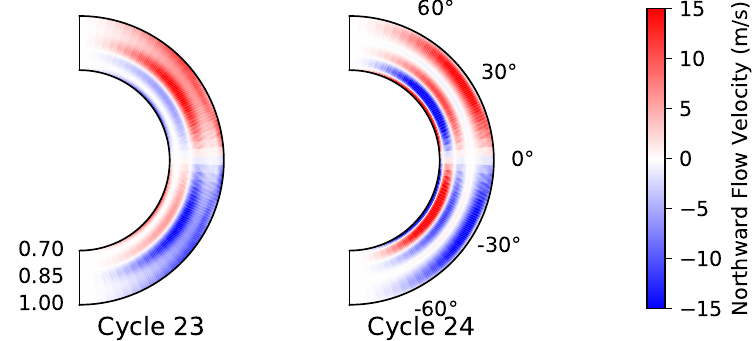}
  \caption{Meridional slice of the inversion domain showing the horizontal flow profiles for each of the examined \cite{2018A&A...619A..99L} data sets.}
  \label{fig:2324polar}
\end{figure}

\begin{figure}
    \centering
    \includegraphics[width=\textwidth]{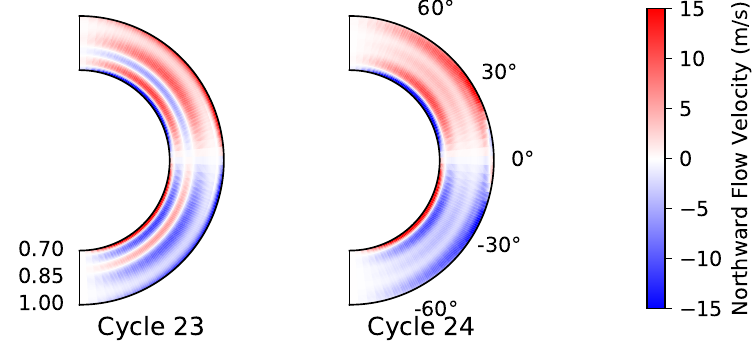}
    \caption{Meridional slice of the inversion domain showing the horizontal flow profiles for each of the examined \cite{2020Sci...368.1469G} data sets using the full error covariance matrix}.
    \label{fig:gizon2324polar}
\end{figure}

The median of the PDF for each parameter is taken as the most likely value. Computing a flow profile from all 12 most likely parameter values results in the "best fit" profile. In Figure\,\ref{fig:2324polar} the resulting best fit flow profiles of cycle 23 and cycle 24 are presented for the \cite{2018A&A...619A..99L} data. Likewise, Figure\,\ref{fig:gizon2324polar} presents the best fit flow profiles for both solar cycles using the \cite{2020Sci...368.1469G} data and the full error covariance matrix. 

Table\,\ref{tab:parranges} shows the complete list of most likely (median) parameter values for each of the six inversions performed in this work, along with bounds corresponding to the 34th and 66th percentile, as computed from the posterior distribution. In addition, the RMS values of the residuals for each inversion are also shown.

\begin{table}[]
  \centering
  \begin{tabular*}{\textwidth}{c@{\extracolsep{\fill}}cccccc}
  \hline
       & Liang 23                              & Liang 24                              & Gizon 23                              & Gizon 24                              & Gizon 23 (full)                        & Gizon 24 (full)                         \\ [0.5ex]
  \hline\hline
  MC\textsubscript{1} & $-1.82\substack{+4.78 \\ -4.81}$      & $-29.75\substack{+4.92 \\ -5.06}$     & $12.99\substack{+3.94 \\ -3.88}$      & $1.51\substack{+3.86 \\ -3.76}$       & $12.36\substack{+4.17 \\ -4.44}$      & $11.75\substack{+4.87 \\ -4.85}$      \\
  MC\textsubscript{2} & $-6.53\substack{+2.71 \\ -2.74}$      & $14.35\substack{+2.75 \\ -2.71}$      & $-4.54\substack{+2.16 \\ -1.97}$      & $2.76\substack{+1.98 \\ -1.79}$       & $-6.05\substack{+2.40 \\ -2.57}$      & $5.76\substack{+2.43 \\ -2.44}$       \\
  MC\textsubscript{3} & $20.55\substack{+2.13 \\ -2.07}$      & $1.53\substack{+1.98 \\ -2.03}$       & $8.83\substack{+1.38 \\ -1.40}$       & $7.04\substack{+1.37 \\ -1.29}$       & $8.80\substack{+1.60 \\ -1.78}$       & $4.70\substack{+2.02 \\ -2.09}$       \\
  MC\textsubscript{4} & $11.51\substack{+1.25 \\ -1.22}$      & $5.73\substack{+1.18 \\ -1.18}$       & $7.39\substack{+0.83 \\ -0.78}$       & $6.18\substack{+0.81 \\ -0.91}$       & $6.89\substack{+0.95 \\ -0.82}$       & $3.26\substack{+1.09 \\ -1.04}$       \\
  MC\textsubscript{5} & $11.58\substack{+3.67 \\ -3.64}$      & $22.80\substack{+3.65 \\ -3.59}$      & $15.82\substack{+2.51 \\ -2.34}$      & $22.00\substack{+2.13 \\ -2.17}$      & $3.25\substack{+2.93 \\ -3.04}$       & $11.26\substack{+3.55 \\ -3.80}$      \\
  MC\textsubscript{6} & $18.12\substack{+7.62 \\ -7.28}$      & $1.32\substack{+6.94 \\ -6.92}$       & $11.57\substack{+5.79 \\ -5.71}$      & $1.67\substack{+4.85 \\ -5.08}$       & $18.12\substack{+6.03 \\ -6.69}$      & $6.64\substack{+7.03 \\ -9.47}$       \\
  MC\textsubscript{7} & $1.24\substack{+0.24 \\ -0.24}$       & $1.45\substack{+0.32 \\ -0.32}$       & $0.68\substack{+0.27 \\ -0.24}$       & $0.65\substack{+0.27 \\ -0.22}$       & $0.25\substack{+0.16 \\ -0.11}$       & $0.59\substack{+0.51 \\ -0.28}$       \\ [0.5ex]
  \hline
  LC\textsubscript{1} & $84.41\substack{+15.37 \\ -13.15}$    & $107.26\substack{+13.60 \\ -12.79}$   & $122.25\substack{+17.66 \\ -17.56}$   & $111.52\substack{+18.45 \\ -17.49}$   & $93.10\substack{+17.26 \\ -17.08}$    & $81.90\substack{+27.38 \\ -21.48}$    \\
  LC\textsubscript{2} & $16.13\substack{+0.95 \\ -0.88}$      & $18.91\substack{+1.40 \\ -1.51}$      & $19.87\substack{+1.41 \\ -1.29}$      & $19.46\substack{+0.82 \\ -1.25}$      & $6.22\substack{+1.75 \\ -1.36}$       & $16.10\substack{+2.58 \\ -11.77}$     \\
  LC\textsubscript{3} & $603.65\substack{+128.75 \\ -142.46}$ & $613.72\substack{+138.09 \\ -154.92}$ & $561.13\substack{+125.34 \\ -125.18}$ & $575.89\substack{+132.71 \\ -160.93}$ & $588.42\substack{+132.34 \\ -155.30}$ & $626.48\substack{+132.53 \\ -129.03}$ \\
  LC\textsubscript{4} & $19.99\substack{+3.46 \\ -3.16}$      & $22.01\substack{+3.88 \\ -3.45}$      & $18.71\substack{+5.69 \\ -3.49}$      & $20.07\substack{+5.75 \\ -3.89}$      & $17.28\substack{+4.50 \\ -3.09}$      & $21.60\substack{+6.68 \\ -4.55}$      \\
  LC\textsubscript{5} & $6.54\substack{+1.19 \\ -1.32}$       & $5.69\substack{+1.38 \\ -1.42}$       & $5.45\substack{+1.31 \\ -1.38}$       & $5.47\substack{+1.46 \\ -1.53}$       & $4.90\substack{+1.68 \\ -2.58}$       & $4.08\substack{+2.14 \\ -2.03}$       \\
  RMS (s)    & $1.40$                                & $1.26$                                & $2.29$                                & $2.05$                                & $2.28$                                & $2.04$    \\
  \hline
  \end{tabular*}
  \caption{Most likely (median) parameter values for all 6 inversions performed in this work. Ranges indicate the 34th and 66th percentile respectively for each parameter shown. Also included is the RMS value of the residuals for each inversion. The "Gizon" column refers to inversions made using only the diagonal of the error covariance matrix, while "Gizon (full)" values are from the inversions which utilize the full noise-covariance  matrix as made available by \cite{2020Sci...368.1469G}. For information on each parameter, see Table\,\ref{tab:paramnames}.}
  \label{tab:parranges}
\end{table}

Resulting flows near the tachocline (from the lower boundary of \(0.7R_\odot\) to about \(0.75R_\odot\)) are not further discussed, as the data are very noisy at large skip distances and there are too few measurements available to adequately probe this region. Previous studies have determined that there is not enough data yet to accurately recover flows at such depths. As noted by \cite{2020Sci...368.1469G}, to recover flows near the tachocline will require somewhere on the order of 100 years of solar observations. On the other hand, \cite{2008ApJ...689L.161B} claims deep return flows may be recovered with as little as one full solar cycle of data. Even when averaging the data to improve signal-to-noise, the 21 years of data utilized in this study proved insufficient for a confident retrieval of deep return flows. The horizontal flow value at the tachocline is also not a free parameter in this particular model, but rather a tuned parameter used to ensure mass conservation. 

First, results of the Bayesian inversions performed on the \cite{2018A&A...619A..99L} data set:
while the flow profile remains more convincingly single-celled in the cycle 23 inversions, the horizontal flow velocity appears to increase in the \(0.85R_\odot\).
The flow structure differs between the two solar cycles studied. From the surface to the tachocline, the cycle 24 results show a poleward flow, followed by a weakening to near zero velocity around \(0.86R_\odot\) (as opposed to a strengthening as observed in cycle 23), and then a strengthening until below \(0.8R_\odot\). Around \(0.78R_\odot\), the flow becomes strongly equatorward. As mentioned previously, below \(0.75R_\odot\) the flow is not constrained by the data and no scientific conclusions can be drawn. Even with the abundance of full-disk solar data available at this time, there is still a need for continued observations if the deep convection zone meridional circulation is to be confidently recovered using current methods.

\begin{figure}
  \centering
  \includegraphics[width=\textwidth]{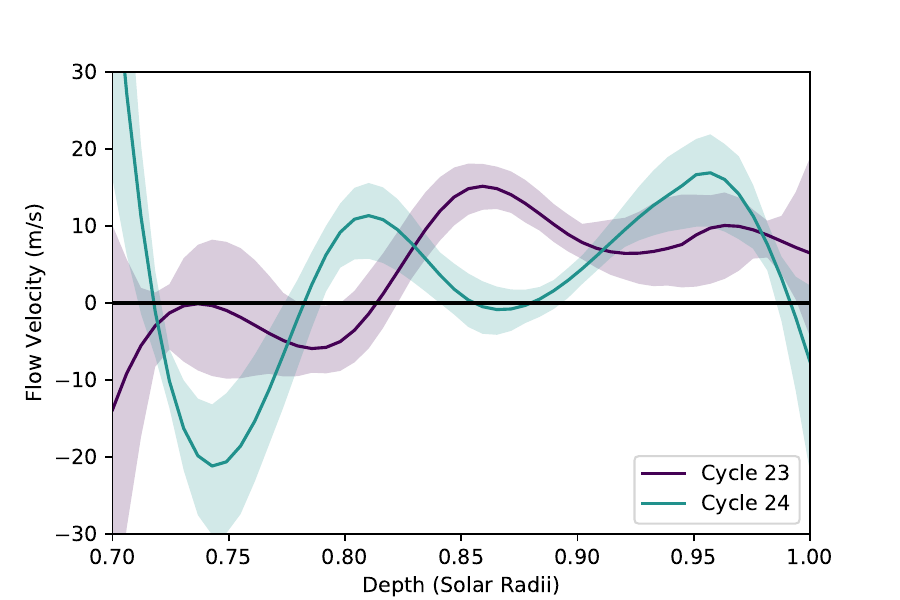}
  \caption{Northward horizontal flow as a function of depth at a latitude of \(30^\circ\) for each of the analyzed Liang} data sets. Shaded regions represent a one-sigma deviation from the most likely flow in each case. The flow presented here is an average over the latitude range \(27^\circ \le \theta \le 33^\circ\).
  \label{fig:cutdepth}
\end{figure}

In order to more easily compare the resulting flow profiles for each component of the \cite{2018A&A...619A..99L} data set, a cut through depth was made centered at \(30^\circ\) latitude and displayed in Figure \ref{fig:cutdepth}. As one can see, the best-fit models over the two studied solar cycles differ most in the \(0.825R_\odot\) to \(0.9R_\odot\) region, with further differences observed at larger depths (though the uncertainty begins to grow rapidly in this regime). 
It is worth noting that the flow profile for the cycle 24 data set at the surface as shown in Figure \ref{fig:cutdepth} shows an equatorward flow at the surface. This is solely due to the latitude selection of \(30^\circ\), which captures the the peak of the LC flow that acts to counter the MC flow in the very near surface region.

The observed decrease in the poleward flow velocity for cycle 24 in the mid-convection zone can, to some extent, be observed directly in the travel-time difference data in \cite{2018A&A...619A..99L}; for skip distances corresponding to depths below \(0.89R_\odot\), the travel-time differences drop to nearly zero. However, this does not explain the additional poleward component that appears centered around \(0.81R_\odot\). The slight decrease in horizontal flow velocity seen in cycle 24 near the surface compared to cycle 23 could be the result of the cycle 24 data set not containing as much of the tail of the solar cycle. It has been found that the meridional circulation becomes weaker during times of sunspot maximum \citep{2021SCPMA..6439601C}, and given the incomplete nature of the cycle 24 data set, more of the data corresponds to times of high activity for this set compared to cycle 23.

\begin{figure}
  \centering
  \includegraphics[width=\textwidth]{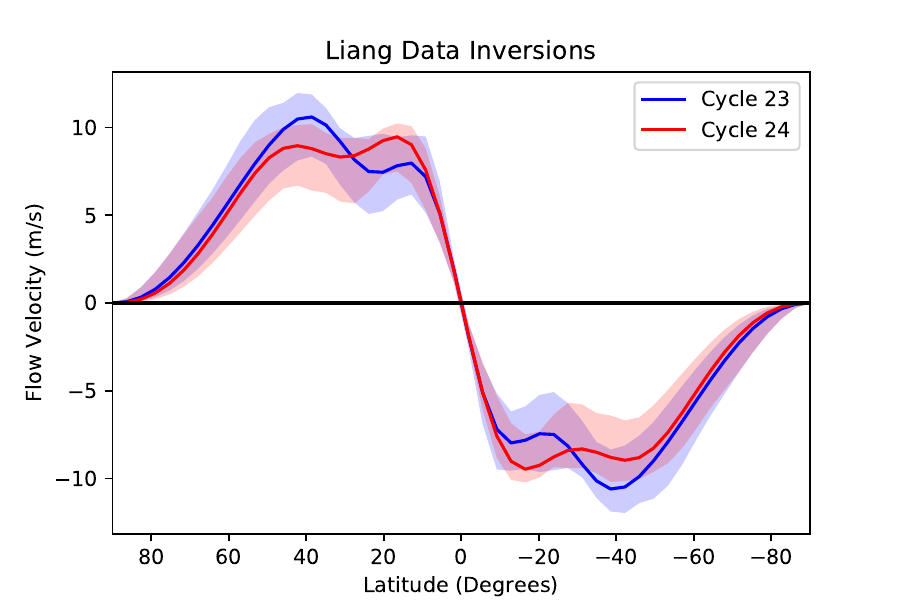}
  \caption{An average of the full flow profile between \(0.95R_\odot\) and \(1.0R_\odot\) across all latitudes for the Liang dataset inversions}. The solid lines represent the most likely model. Shaded regions represent one sigma deviations from the most likely solution. Note the localized reduction of velocity occurring around the active latitudes as a result of LC flows.
  \label{fig:shallowdepth}
\end{figure}

Compared to the results of \cite{2020Sci...368.1469G}, the flow velocities found in this study differ in a couple of key ways. At the same latitude of \(30^\circ\), \cite{2020Sci...368.1469G} finds flow profiles for both cycles that decrease from \(0.95R_\odot\) to \(0.8R_\odot\) far more monotonically than in this study. While the cycle 24 results for \cite{2020Sci...368.1469G} are indeed consistent with zero flow in the southern hemisphere between roughly \(0.88R_\odot\) and \(0.85R_\odot\), there is no indication of any poleward flow deeper than this region as is found in this investigation (Figure \ref{fig:cutdepth}). In general, there is better morphological agreement between \(0.95R_\odot\) and \(0.75R_\odot\) for cycle 23, though this investigation finds higher poleward flow velocities. The depth of the final zero-crossing indicating an equatorward flow occurs in this study between \(0.82R_\odot\) and \(0.78R_\odot\) for cycle 23 and cycle 24 respectively, which is broadly consistent between both studies. An additional cut can be made at a given depth to compare how the horizontal flow profile changes with latitude for each solar cycle. Such results are shown in Figure \ref{fig:shallowdepth}. Compared to recent results from \cite{2020Sci...368.1469G}, this work finds weaker flows in general. However, the decrease in flow velocity at about \(30^\circ\) latitude due to an LC component and peak velocities of around 10 meters per second are seen in both works. Note that in this investigation, the flow profiles produced are antisymmetric about the equator; as such, differences between the hemispheres are not detectable, though such differences are seen in \cite{2020Sci...368.1469G}. As our uncertainties at the surface are very large due to few short-distance travel times, Figure\,\ref{fig:shallowdepth} shows an average of the horizontal flow profile between \(0.95R_\odot\) and \(1.0R_\odot\), while the corresponding panel C in Figure\,3 of \cite{2020Sci...368.1469G} only shows the surface measurement. The shortest skip distance measurement used in this analysis has a lower turning point of \(0.963R_\odot\), and therefore flows computed nearer to the surface than this are effectively extrapolations and should be considered skeptically.

As for other results that differ from this work, \cite{2015ApJ...805..133J} studied two years of GONG data obtained between 2004 and 2012 and found no evidence of a double-celled meridional flow profile. However, those data correspond to the tail end of cycle 23 and the beginning of cycle 24, and are not necessarily directly comparable to the cycle 24 results found in this work. On the other hand, \cite{2015ApJ...813..114R} found a distinctly single-celled profile using four years of HMI data, all of which falls within the cycle 24 data range used in this investigation. \cite{2015ApJ...813..114R} also found higher flow velocities near the surface than was found in this study. Despite these differences, some qualitative similarities exist. As can be seen in the right panel of \cite{2015ApJ...813..114R}'s Figure\,3, the horizontal flow velocities exhibit a weakening with a minimum near \(0.9R_\odot\) before strengthening to a peak again at \(0.83R_\odot\). While the velocities are more consistent with this study's cycle 23 results, there is still evidence for a broad weakening in the horizontal flow at mid-depths. The results of \cite{2018ApJ...863...39M} also show quite different flow profiles: their Figure\,8 shows a broad weakening in the horizontal flow profile that occurs near \(0.85R_\odot\). This is quite a bit deeper than \cite{2015ApJ...813..114R} but more consistent with this study's cycle 24. Below this, the horizontal flow strengthens, but only very slightly - while this study's flow recovers to near-surface values, \cite{2018ApJ...863...39M}'s flow only recovers to less than half of their peak velocity.

Previous investigations have also found evidence of double-celled meridional flow profiles. \cite{2017ApJ...849..144C} used seven years of HMI data from cycle 24 and found a mid-convection zone equatorward flow region between roughly \(0.90R_\odot\) and \(0.82R_\odot\). While our results are only consistent with a weakening of the poleward flow rather than a full detection of a return flow in this region, the two flow profiles are qualitatively similar. \cite{2017ApJ...849..144C} find a poleward flow component deeper and with a stronger latitudinal dependence compared to what was found in this study. \cite{2013ApJ...774L..29Z} find a similarly consistent double-celled flow profile using two years of HMI data which also correspond to cycle 24. The mid-convection zone equatorward flow is once again broader but still consistent with the weakening of the poleward flow flow region found in this study.

\begin{figure}
    \centering
    \includegraphics[width=\textwidth]{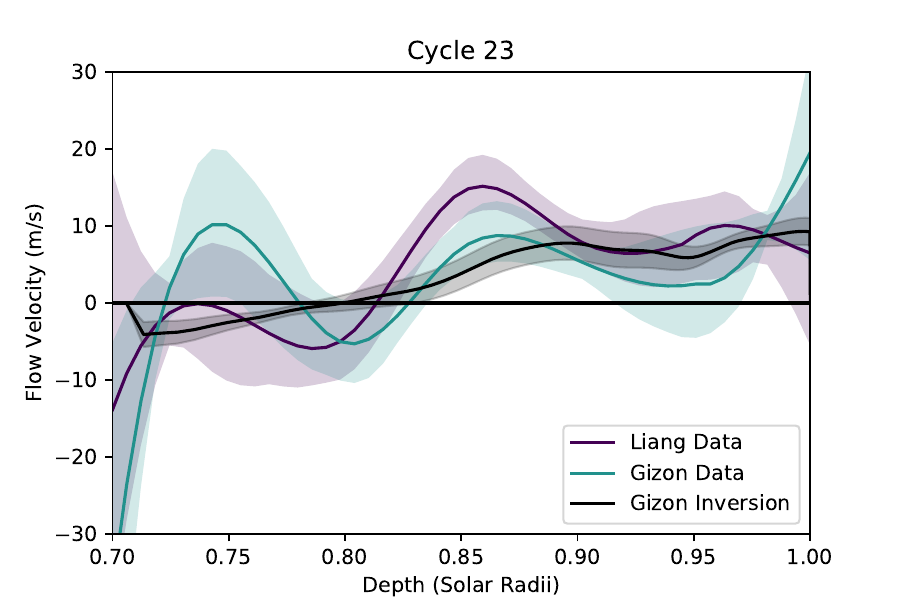}
    \caption{A comparison of the horizontal flow profile outputs for solar cycle 23 using \cite{2018A&A...619A..99L} data (purple) and \cite{2020Sci...368.1469G} data (cyan). The black line represents the inversion performed by the authors of \cite{2020Sci...368.1469G} for the same cycle. Shaded regions represent a one-sigma deviation from the most likely flow profile in each case. The flow presented here is an average over the latitude range \(27^\circ \le \theta \le 33^\circ\).}
    \label{fig:gizonliang23}
\end{figure}

\begin{figure}
    \centering
    \includegraphics[width=\textwidth]{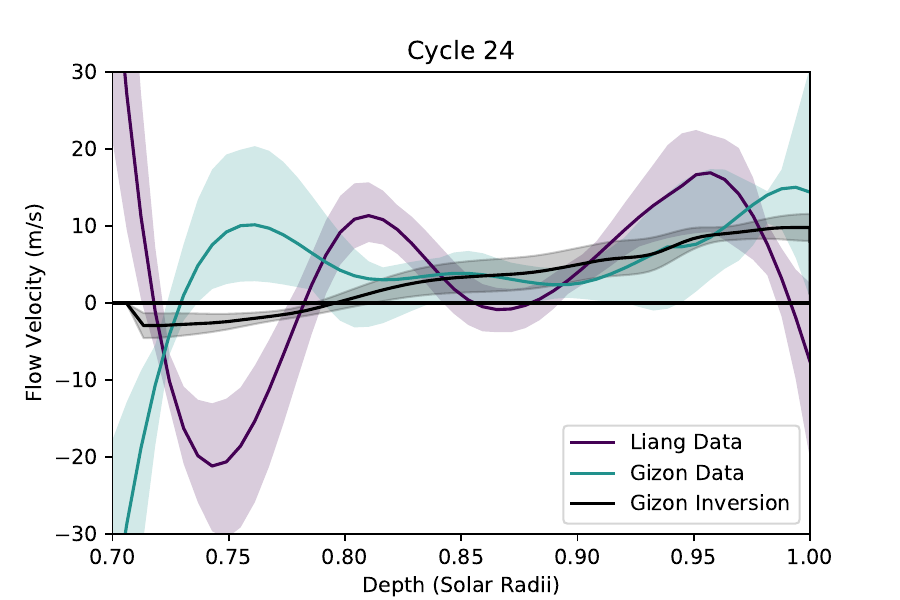}
    \caption{A comparison of the horizontal flow profile outputs for solar cycle 24 using \cite{2018A&A...619A..99L} data (purple) and \cite{2020Sci...368.1469G} data (cyan). The black line represents the inversion performed by the authors of \cite{2020Sci...368.1469G} for the same cycle. Shaded regions represent a one-sigma deviation from the most likely flow profile in each case. The flow presented here is an average over the latitude range \(27^\circ \le \theta \le 33^\circ\).}
    \label{fig:gizonliang24}
\end{figure}

\begin{figure}
  \centering
  \includegraphics[width=\textwidth]{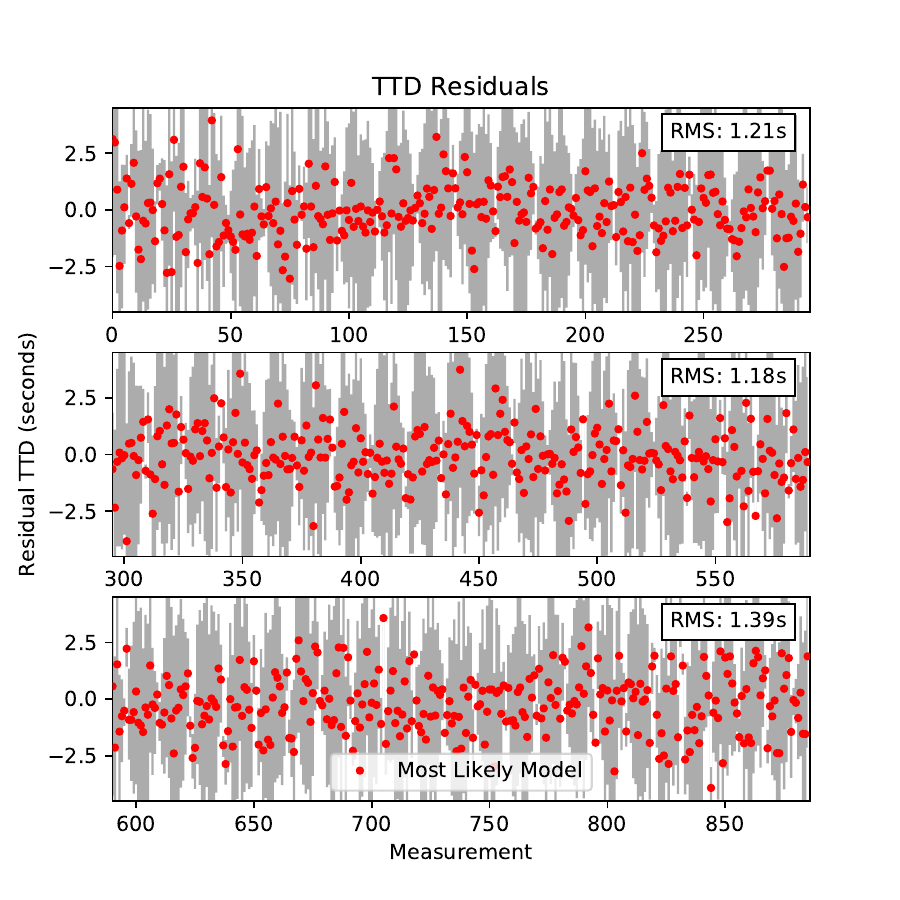}
  \caption{Residuals for the most likely model compared to the input data for the Liang cycle 24 data set. The gray bars represent the range of residuals from 50 random draws from the priors in order to illustrate the extent to which the model can differ from the input data. Data points are plotted for every 10th point so as to not clutter the plot. RMS values for the most likely model in each panel are labeled. The RMS for the full data set is 1.26 seconds.}
  \label{fig:robustness-24}
\end{figure}

In order to understand our results more completely, the residuals between the inversion results in data space for cycle 24 and the input data set are shown in Figure \ref{fig:robustness-24}. Such a plot is an attempt to display the robustness of the model in fitting the data by including not only the residual travel times computed using the most likely flow profile, but also the residual travel times computed from a large number of random models sampled from the prior distribution.

The region in which the cycle 23 and 24 inversions differ the most (around a depth of \(0.85_\odot\)) is captured in the middle panel of Figure\,\ref{fig:robustness-24}. This panel also has the lowest RMS, and one can see that the MC parameters that control the horizontal flow velocity near these depths (MC\textsubscript{3} and MC\textsubscript{4}) tend to be more well-constrained (see Table\,\ref{tab:parranges}). This is also visible in Figure\,\ref{fig:cutdepth} and similar figures, where the one-sigma range shrinks below \(0.9_\odot\).
This may be explained due to the lack of near-surface sampling and the increased noise at large depths leading to a "sweet spot" of adequate coverage and reasonable noise in the mid-depths.

Random draws from the prior distribution show that the range of travel time differences that can be obtained using this model vary quite significantly from the value for the most likely model in the top panel of Figure\,\ref{fig:robustness-24}. The gray bars show the travel time difference residuals for the random samples of the prior distribution. The range of the residuals indicates that the model is capable of producing flows that yield travel time differences up to 5 or more seconds greater than or less than those found in the data. This indicates that even at large depths, the model is capable of reproducing flows that match the data.

\begin{figure}
  \centering
  \includegraphics[width=\textwidth]{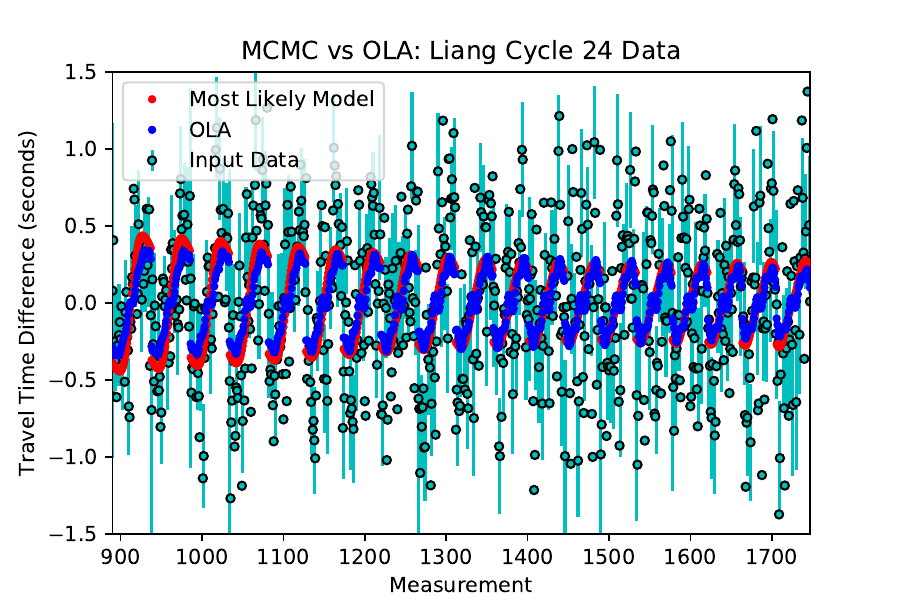}
  \caption{A comparison of the best-fitting model from the Bayesian inversion in data-space compared to a SOLA inversion for cycle 24, restricted to the region in which the most likely flows deviate the most between cycles - from \(0.825R_\odot\) to \(0.9R_\odot\). The red points show the travel-time difference measurements generated from the most likely flow parameters, while the blue points show travel-time difference measurements from a SOLA inversion on the same data set. The teal points represent the data used for both cases. Error bars are included for every fourth data point so as to not clutter the plot.}
  \label{fig:liangmcmcola24}
\end{figure}

\begin{figure}[t]
  \centering
  \includegraphics[width=0.7\textwidth]{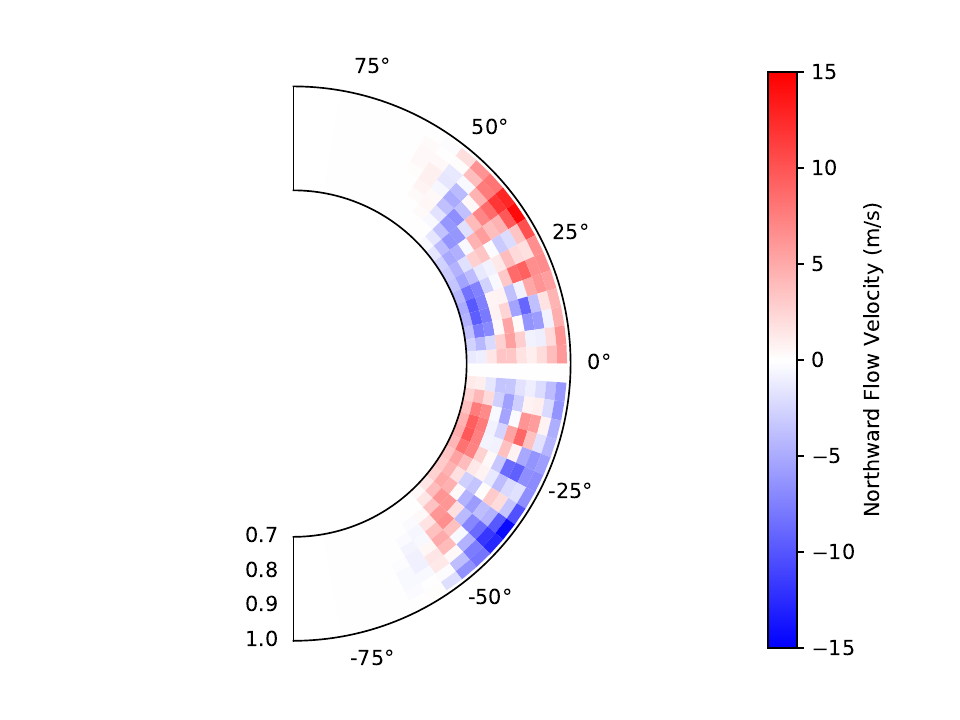}
  \caption{Meridional slice of the inversion domain showing the horizontal flow profile of an SOLA inversion performed on the \cite{2018A&A...619A..99L} cycle 24 data set.}
  \label{fig:olapolar}
\end{figure}

In an effort to compare the Bayesian results to non-Bayesian inversions on the same data, SOLA inversions were carried out on the \cite{2018A&A...619A..99L} data as well. Figure\,\ref{fig:liangmcmcola24} shows a comparison between the best-fit model from the Bayesian inversions for cycle 24 compared to an OLA inversion. The SOLA inversion was performed after averaging the data set by a factor of four using a weighted averaging technique. In order to compare directly, the MCMC inversion was also performed on this averaged data set for this comparison. As noted in \cite{2020SoPh..295..137J}, OLA inversions systematically underestimate the amplitude of the flow. The impact of these underestimated flows can be observed in the consistently lower travel-time differences of the OLA points compared to the Bayesian inversion's results. Interestingly, while the travel-time differences are quite similar between the two, the flow profiles differ more substantially. Figure\,\ref{fig:olapolar} shows a meridional slice through the flow results of the OLA inversion (compare to the right panel of Figure\,\ref{fig:2324polar}). The difference in flow profile despite the similarity in travel-time differences point to either a degeneracy in well-fitting flow profiles (i.e., multiple flow profiles exist which yield appropriate travel-time differences) or a lack of sensitivity of the model used. To rule out a lack of sensitivity in the model, tests were carried out by forward modeling travel-time differences from a series of disparate artificial flow profiles. The results of these tests showed conclusively that the model is indeed capable of producing a wide range of travel-time differences for different flow profiles. No significant difference was found in the quality of the fit between the Bayesian inversion's best-fit profile and the OLA inversion.

\begin{sidewaysfigure}
  \centering
  \includegraphics[width=\textwidth]{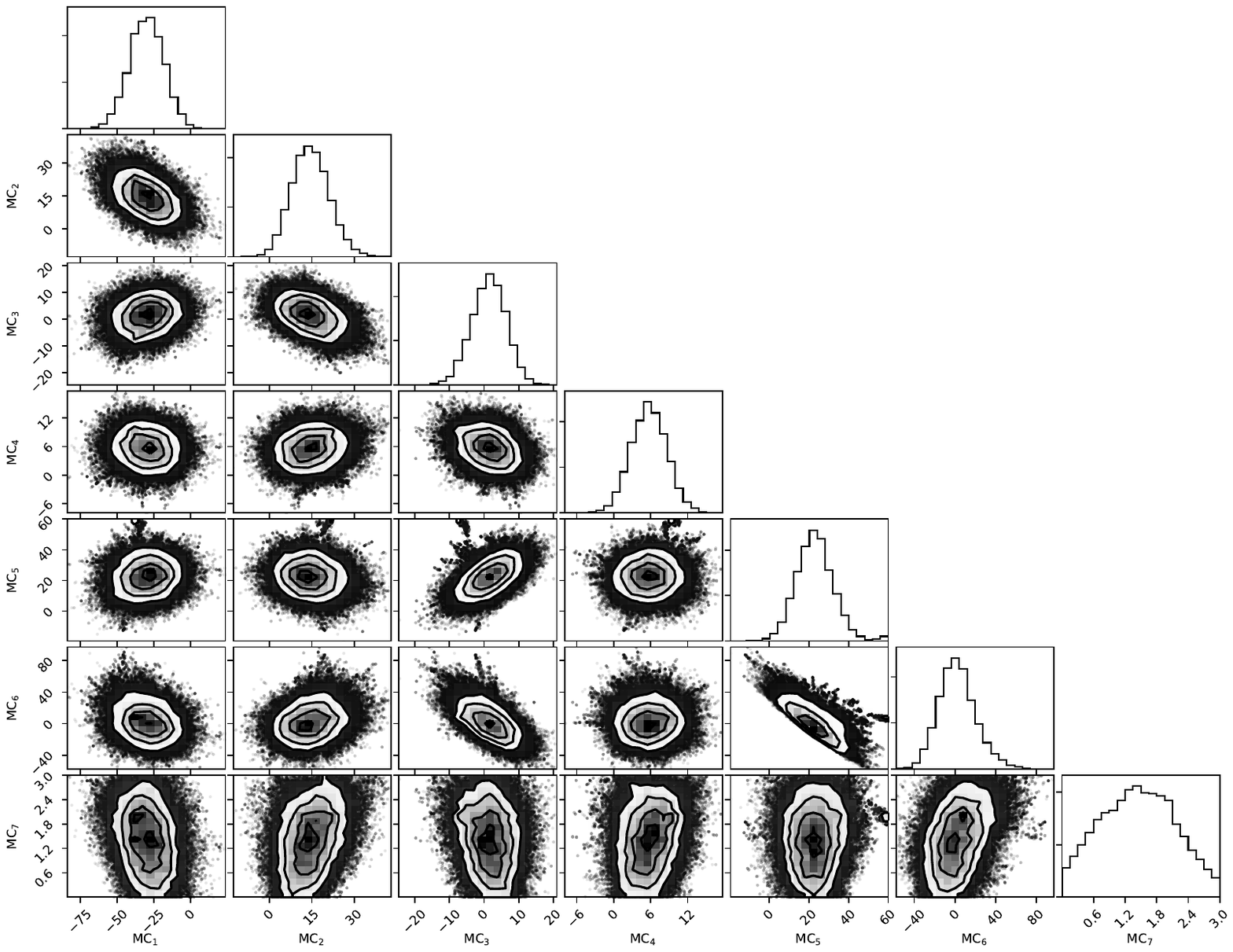}
  \caption{Corner plot of MC flow parameters for the cycle 24 data set. Refer to Table \ref{tab:paramnames} for parameter details. Contours represent 0.5, 1.0, 1.5, and 2.0 sigma levels.}
  \label{fig:24mc}
\end{sidewaysfigure}

\begin{sidewaysfigure}
  \centering
  \includegraphics[width=\textwidth]{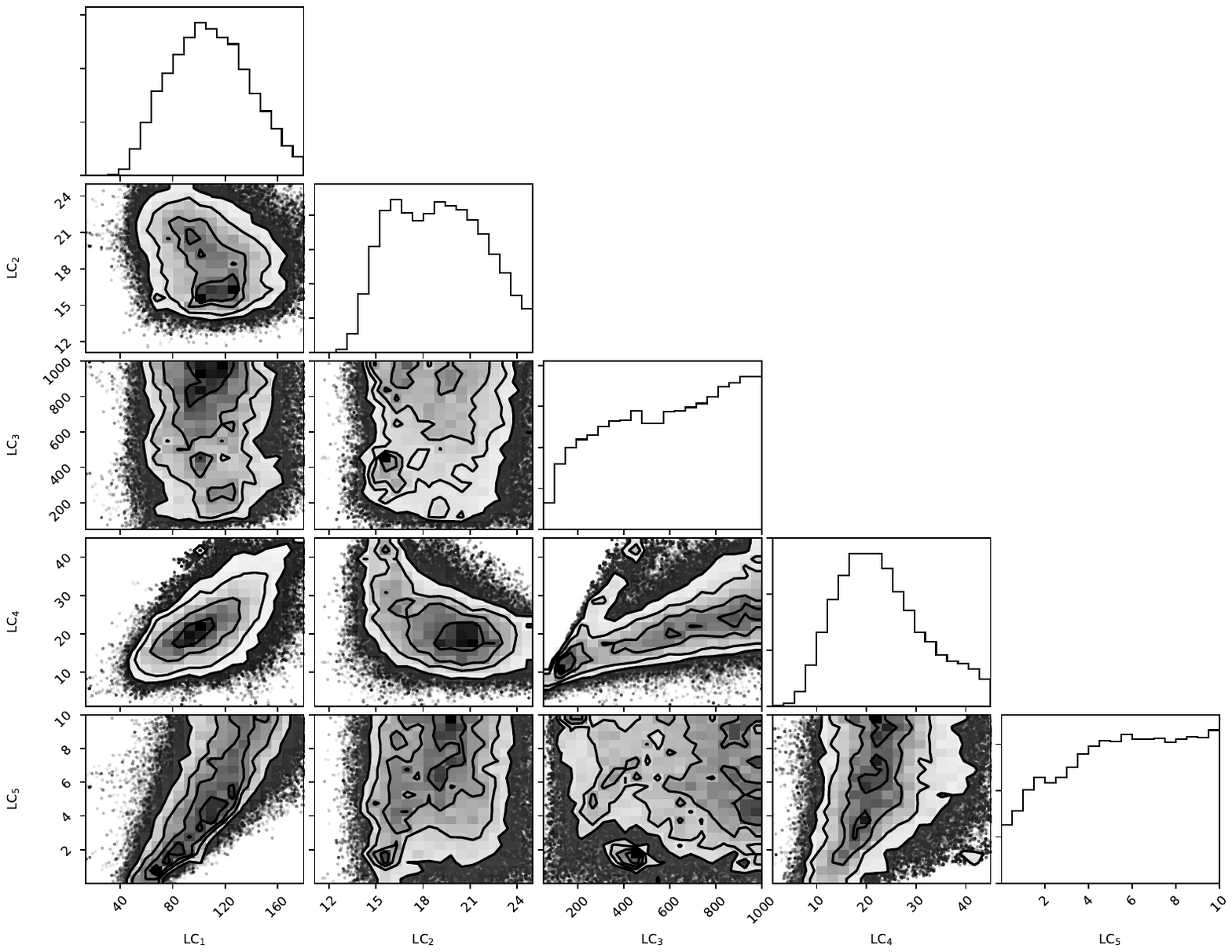}
  \caption{Corner plot of LC flow parameters for the cycle 24 data set. Refer to Table \ref{tab:paramnames} for parameter details. Contours represent 0.5, 1.0, 1.5, and 2.0 sigma levels.}
  \label{fig:24lc}
\end{sidewaysfigure}

Figures\,\ref{fig:24mc} and \ref{fig:24lc} show the ``corner'' plots of the \cite{2018A&A...619A..99L} data set's cycle 24, where the marginalized posterior distribution for each parameter can be compared against each other parameter to visualize the correlation between parameters. In addition, following one parameter's column or row to the end will bring one to the histogram of the posterior for that parameter. Parameters with more well-defined peaks in their posterior distribution (such as MC\(_2\) in Figure\,\ref{fig:24mc}) have smaller variance, and departing from the peak lowers the quality of the fit significantly. On the other hand, parameters with a flatter posterior (such as LC\(_5\) in Figure\,\ref{fig:24lc}) have larger variance and are less constrained by the data. Overall, there are no strong correlations between most of the model parameters. Additionally, there is no presence of a double-peak for any MC parameter, indicating that there is little chance for ambiguity between a single- and double-celled best-fit profile for cycle 24.

Finally, to more directly investigate the differences between the Bayesian inversions performed on the \cite{2018A&A...619A..99L} data and the results of \cite{2020Sci...368.1469G}, Bayesian inversions were also performed using the same data \cite{2020Sci...368.1469G} used. While inversions were performed using both the full covariance matrix and the diagonal of the error matrix (equivalent to the errors used in the inversions on the \cite{2018A&A...619A..99L} data set), the results that best matched the data were the runs using the full covariance matrix. Differences between the two runs were largely insignificant. As such, the results shown in the following figures are from the full covariance inversions. Figure\,\ref{fig:gizon2324polar} shows a meridional slice through the two best-fit flow profiles for cycle 23 and cycle 24 respectively. The profiles obtained using this data set differed significantly from the previous result, and are much more in line with the authors' findings. For comparison, Figures\,\ref{fig:gizonliang23} and \ref{fig:gizonliang24} show the same horizontal flow as a function of depth as in Figure\,\ref{fig:cutdepth}, but with the results of the Bayesian inversion on both sets of data for each solar cycle compared to \cite{2020Sci...368.1469G}'s own inversion.

For cycle 23, a similar qualitative structure can be seen between the Bayesian inversion of the \cite{2018A&A...619A..99L} and \cite{2020Sci...368.1469G} flow profiles between \(0.8R_\odot\) and \(0.97R_\odot\). However, within this region, the Bayesian inversion on the \cite{2020Sci...368.1469G} data much more closely matches the authors' own inversion than that of the Bayesian inversion performed on the \cite{2018A&A...619A..99L} data: the two results are consistent to within one sigma from the surface to around \(0.76R_\odot\). Cycle 24 shows more consistency between the Bayesian inversions performed on both data sets down to a depth of \(0.8R_\odot\) or so, but only the Bayesian inversion using the \cite{2020Sci...368.1469G} data stayed within one sigma of the authors' inversion once again. Even stronger bizarre behavior below \(0.8R_\odot\) is seen in cycle 24, with the \cite{2020Sci...368.1469G} data set inversion jumping to higher positive flow velocities followed by a sharp return to large, negative velocities near the tachocline. The inversion results on the \cite{2018A&A...619A..99L} data set show nearly inverted behavior in this region. While the flows do differ slightly above this depth, no indication of any significant difference between the two solar cycles near \(0.86R_\odot\) is seen with this data set. Overall, as found by \cite{2020Sci...368.1469G}, the Bayesian inversions on their data indicate the presence of a single-celled flow profile in both solar cycles.

\section{Summary} \label{sec:summary}

Using up to twenty-three years of HMI, MDI, and GONG data and a meridional circulation model from \cite{2018A&A...619A..99L}, this work developed a Bayesian Markov chain Monte Carlo method to explore a highly-dimensional parameter space and determine the meridional circulation. The model is comprised of twelve free parameters; seven for the large-scale meridional circulation (MC) profile and five for the shallow, local cellular (LC) inflow observed near active regions.

These data consist of two distinct sets of travel-time differences: data from \cite{2018A&A...619A..99L} and data from \cite{2020Sci...368.1469G} - each with two components corresponding to cycle 23 and cycle 24. Both sets of data were averaged antisymmetrically across the equator to increase the signal-to-noise and an iterative sigma cut was employed to eliminate the most egregious outliers. The inversion samples several hundred thousand flow models for each data set.

After validating the inversion with various tests with artificial data, we find that the when utilizing the \cite{2018A&A...619A..99L} data set, meridional circulation in the Sun shows some difference between the last two solar cycles (Figure\,\ref{fig:2324polar}). Cycle 23 shows a large, single-celled profile with flow velocities of around $15\,{\rm m\,s^{-1}}$ throughout much of the convection zone, and a weaker return flow near the tachocline as expected and found in other meridional circulation studies \citep[][among others]{2020Sci...368.1469G,2020ASSP...57..107R}. However, cycle 24 shows a different flow profile with a weaker poleward flow at the surface, dropping to values consistent with zero around \(0.86R_\odot\), returning poleward again and then finally returning equatorward near the tachocline. Such a flow profile is generally consistent with some previous studies of HMI data for cycle 24 (\cite{2013ApJ...774L..29Z}, \cite{2017ApJ...849..144C}), though there are works showing a more single-celled flow profile for the same time period \citep{2015ApJ...813..114R,2020Sci...368.1469G}. 

A second set of data were used to compare the results of this Bayesian model directly to inversions made without Bayesian techniques. This data set was investigated in detail by \cite{2020Sci...368.1469G}. Compared with their results, the Bayesian approach yielded somewhat different flow profiles. The non-Bayesian approach seems to prefer a smoother, less variable flow profile than the Bayesian method used in this work, largely due to limitations in the model.

Comparing the results of the Bayesian approach using the \cite{2018A&A...619A..99L} data versus the same approach applied to the \cite{2020Sci...368.1469G} over the same time frame illustrates some of the shortcomings of the model used. Near-surface flows produced by this model show very large uncertainties, while deep measurements exhibit unusual or even unphysical behavior across both data sets utilized.

Uncertainty in the flow profiles increases drastically near the tachocline (Figure \ref{fig:cutdepth}), and as such flow velocities below around \(0.75R_\odot\) are unreliable. Despite the great deal of available data for such travel-time difference measurements, confidently recovering the flow profile near the tachocline will take on the order of 100 years of data \citep{2020Sci...368.1469G}. While the model used may have some unfortunate peculiarities, the underlying Bayesian approach is functional and yields similar quality fits to the data as OLA inversions do (Figure\,\ref{fig:liangmcmcola24}). Further study is required to investigate whether the Bayesian approach can provide a more significant improvement over traditional methods by using a more suitable model of solar meridional circulation. 

\begin{acknowledgments}
This work is supported by NASA under award number 80NSSC18K0672.
\end{acknowledgments}

-----
\clearpage
\bibliographystyle{aasjournal}
\bibliography{references}
\end{document}